\documentclass{phc-proc4-auth}

 \usepackage{graphicx}

\usepackage{amssymb}

\begin{document}

\begin{frontmatter}



\title{Fermi surface topology and vortex state in MgB$_2$}


\author{T. Dahm \corauthref{cor1}},
\author{S. Graser, and N. Schopohl}
\corauth[cor1]{Corresponding author. e-mail: dahm@uni-tuebingen.de}
\address{Institut f\"ur Theoretische Physik, 
         Universit\"at T\"ubingen, 
         Auf der Morgenstelle 14, D-72076 T\"ubingen, 
         Germany}

\begin{abstract}
Based on a detailed modeling of the Fermi surface topology of
MgB$_2$ we calculated the anisotropy of the upper critical field
B$_{c2}$ within the two gap model. The $\sigma$-band is modeled
as a distorted cylinder and the $\pi$-band as a half-torus, with
parameters determined from bandstructure calculations. Our results 
show that the unusual strong temperature dependence of the
B$_{c2}$ anisotropy, that has been observed recently,
can be understood due to the small c-axis dispersion of the 
cylindrical Fermi surface sheets and the small interband 
pairing interaction as obtained from bandstructure calculations.
We calculate the magnetic field
dependence of the density of states within the vortex state
for field in c-axis direction and compare with recent measurements
of the specific heat on MgB$_2$ single crystals.
\end{abstract}

\end{frontmatter}


It now appears widely accepted that the superconducting state of
MgB$_2$ has to be described by two different $s$-wave gaps
\cite{Canfield}. This also implies unusual behavior in its
vortex state, which has been observed recently. The anisotropy
of the upper critical field $B_{c2}$ displays an anomalous strong
temperature dependence. An upward curvature of $B_{c2}$ close to $T_c$ is
observed, when the field is directed in ab-plane direction \cite{Angst,Lyard}.
The specific heat and thermal conductivity show a rapid increase
as a function of magnetic field \cite{Bouquet,Sologubenko} and
the size of the vortex core and local density of states show
unusual behavior as well \cite{Eskildsen}. While it seems
reasonable that these observations are related to the two-gap
nature of MgB$_2$, a detailed understanding of all of these phenomena
within a single model is still lacking.
We have shown recently that the unusual behavior of the upper critical
field can be understood, if the topology of the Fermi surface and the
weak interband pairing interaction is taken into account \cite{Dahm}.
Within our model the $\sigma$-band is described as a distorted cylinder
and the $\pi$-band as a half-torus, in close agreement with bandstructure
calculations.
\begin{figure}[t]
  \begin{center}
    \includegraphics[width=0.75\columnwidth,angle=270]{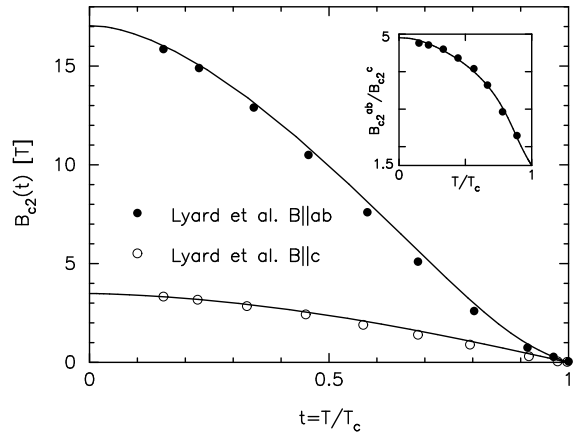}
    \caption{Upper critical field as a function of temperature for field in
     c-axis and ab-plane direction. Comparison with Lyard et al. \cite{Lyard}
     (symbols).
     The inset shows the strong temperature dependence of the anisotropy
     ratio $B_{c2}^{ab}/B_{c2}^c$.
    \label{fig1} }
  \end{center}
\end{figure} 
In Fig.~\ref{fig1} we present a direct comparison of the $B_{c2}$ data
by Lyard et al \cite{Lyard} with our calculation for interband pairing
strength $\eta=0.121$ and c-axis dispersion $\epsilon_c=0.182$ of the 
cylindrical Fermi surface sheet as described in Ref. \cite{Dahm}. For field 
in c-axis direction the temperature dependence follows very much a
conventional one-band temperature dependence \cite{Posa}. 
However, a clear upward curvature close to $T_c$
is obtained for field in ab-plane direction. This results
from a strong temperature dependence of the distortion of the
vortex lattice, which at low temperatures is dominated by the
$\sigma$-band, while close to $T_c$ both bands contribute
equally \cite{Dahm}.

\begin{figure}[t]
  \begin{center}
    \includegraphics[width=0.75\columnwidth,angle=270]{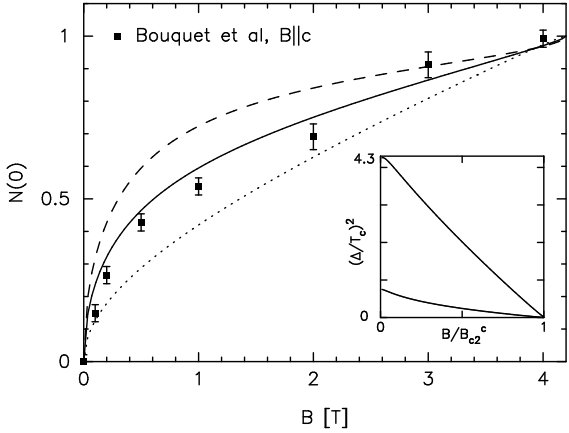}
    \caption{
     Zero energy density of states as a function of magnetic field in 
     $c$-axis direction (solid line) and comparison with
     results from specific heat measurements by Bouquet et al. \cite{Bouquet}
     (symbols) on MgB$_2$ single crystals taking $B_{c2}^c=4.2$~T.
     The inset shows the field dependence of $(\Delta/T_c)^2$ for the two
     gaps.
     \label{fig2}}
  \end{center}
\end{figure} 

Since our model offers a good description of $B_{c2}$ in MgB$_2$
we extended our calculations now to magnetic fields below $B_{c2}$ 
using the analytical method described in Ref. \cite{Graser}
for the calculation of the density of states within the vortex state.
Within this method the field dependence of the partial density of states 
of each band at the Fermi level for field in c-axis direction is given by
\begin{equation}
N_\alpha \left(\omega=0, B \right) = \Bigg\langle 
\frac{N_\alpha(0)}{\sqrt{ 
1 + \frac{4 \Phi_0 \Delta_\alpha^2(B)}{\pi \hbar^2 B v_{F,\alpha,ab}^2}
}}
\Bigg\rangle_\alpha
\label{eq1}
\end{equation}
Here, $v_{F,\alpha,ab}$ is the ab-plane component of the Fermi velocity
of band $\alpha \in \{ \sigma, \pi\}$. $N_\alpha(0)$ is the normal state
density of states, $\Phi_0$ the flux quantum, and $\langle \cdots \rangle_\alpha$
denotes a Fermi surface average over band $\alpha$. For the cylindrical and 
torus Fermi surfaces 
considered here these averages can be done analytically. The field dependence
of the gap $\Delta_\alpha(B)$ is obtained from a solution of
the two by two gap equation in the vortex state \cite{Graser}.

In Fig.~\ref{fig2} we show the magnetic field dependence of the total density of
states at the Fermi level (solid line) for the same set of parameters as in 
Fig.~\ref{fig1} showing fair agreement with the specific heat data of
Bouquet et al. \cite{Bouquet}.
While the contribution from the large gap ($\sigma$ band, dotted line) to a 
good approximation follows the behavior of a single band $s$-wave superconductor, 
the contribution from the small gap ($\pi$ band, dashed line) displays 
a rapid increase. The inset shows the field dependence of the 
square of the two gaps. It shows that the square of
the large gap to a very good approximation is a linear function of the
magnetic field, while the small gap shows some reduction at higher fields.
If we nevertheless approximate the squares of the gaps as linear functions
in $B/B_{c2}^c$ we find the following useful analytical expressions:
\[
N_\sigma(B) \approx \left[ 1 + 0.36 \frac{\Delta^2_\sigma(0)}{T_c^2}
\left( \frac{B_{c2}^c}{B} -1 \right) \right]^{-1/2}
\]
and
\[
N_\pi(B)=\frac{2}{\pi-2 \kappa} \left[ \arcsin k -
\frac{\kappa}{k} E(k) - \kappa \frac{1-k^2}{k} K(k)\right]
\]
with
$
k(B) \approx \left[ 1 + 0.36 \frac{\Delta^2_\pi(0)}{T_c^2}
\frac{v^2_{F,\sigma}}{v^2_{F,\pi}}
\left( \frac{B_{c2}^c}{B} -1 \right) \right]^{-1/2}
$. Here, $E(k)$ and $K(k)$ are the complete elliptical integrals and
$\kappa=0.25$ the ratio of the two radii of the torus Fermi surface
as defined in Ref. \cite{Dahm}.
At low magnetic fields from Eq.~(\ref{eq1}) we can obtain the following
limiting expressions:
$N_\sigma(B) \approx 1.67 \frac{T_c}{\Delta_\sigma(0)} \sqrt{\frac{B}{B_{c2}^c}}$
and
$N_\pi(B) \approx 1.01 \frac{T_c}{\Delta_\pi(0)} \frac{v_{F,\pi}}{v_{F,\sigma}}
\sqrt{\frac{B}{B_{c2}^c}}$.
These expressions show that the rapid increase of the density of states
of the $\pi$-band is due to two reasons: the small value of $\Delta_\pi(0)/T_c$
and the higher in-plane Fermi velocity of the $\pi$-band and thus has a
different physical origin than the
$d$-wave behavior known as Volovik's law \cite{Volovik}. 
We note, however, that these analytical expressions are only valid for
field directed into c-axis direction. For other field directions the
distortion of the vortex lattice has to be taken into account and
we find important modifications. Details will be published elsewhere \cite{DGS}.

\end{document}